\documentclass[12pt,A14]{JHEP3}
\pdfoutput=1
\usepackage{amssymb, amsmath, amsopn, amsthm}

\usepackage{epsfig}

\usepackage{graphics}

\def\makeatletter{\catcode`\@=11}
\makeatletter
\def\mathbox#1{\hbox{$\m@th#1$}}%
\def\math@ccstyles#1#2#3#4#5#6#7{{\leavevmode
      \setbox0\mathbox{#6#7}%
      \setbox2\mathbox{#4#5}%
      \dimen@ #3%
      \baselineskip\z@\lineskiplimit#1\lineskip\z@
      \vbox{\ialign{##\crcr
             \hfil \kern #2\box2 \hfil\crcr
             \noalign{\kern\dimen@}%
             \hfil\box0\hfil\crcr}}}}
\def\mathaccstyles{\math@ccstyles\maxdimen}
\def\maththroughstyles{\math@ccstyles{-\maxdimen}}
\def\unity%
 {\maththroughstyles{.45\ht0}\z@\displaystyle {\mathchar"006C}\displaystyle 1}

\title{Holography Of Charged Dilaton Black Holes}

\author{Kevin Goldstein$^{1}$, Shamit Kachru$^{2}$\footnote{On leave of absence from Department of Physics and SLAC, Stanford University.},~Shiroman Prakash$^3$~and Sandip P. Trivedi$^{3}$

  ~\\
  $^1$National Institute for Theoretical Physics (NITHeP), \\
  School of Physics and Centre for Theoretical Physics, \\
  University of the Witwatersrand, \\
  WITS 2050 \\
  Johannesburg, South Africa \\
  \vspace{0.1cm}

$^2$Kavli Institute for Theoretical Physics and Department of Physics\\
 University of California \\
Santa Barbara, CA 93106\\
 \vspace{0.1cm}

$^3$Tata Institute for Fundamental Research \\
Mumbai 400005, India\\

\vspace{0.1cm}

\email{Email: kevin.goldstein@wits.ac.za, skachru@kitp.ucsb.edu, shiroman@gmail.com, trivedi.sp@gmail.com}\\

}

\abstract{ We study charged dilaton black branes in $AdS_4$.
  Our system involves a dilaton $\phi$ coupled to a Maxwell field $F_{\mu\nu}$ with dilaton-dependent gauge coupling,
  ${1\over g^2} = f^2(\phi)$.  
  First, we find the solutions for extremal and near extremal branes through a combination
  of analytical and numerical techniques.  The near horizon geometries in the simplest 
  cases, where $f(\phi) = e^{\alpha\phi}$, are Lifshitz-like,
  with a dynamical exponent $z$ determined by $\alpha$.  
  The black hole thermodynamics varies in an interesting way with $\alpha$, but in all cases
  the entropy is vanishing and the specific heat is positive for the near extremal solutions.
  We then compute conductivity in these backgrounds.  We find that somewhat surprisingly, the 
  AC conductivity vanishes like $\omega^2$ at $T=0$ independent of $\alpha$.  
  We also explore the charged black brane physics of several other classes of gauge-coupling functions
  $f(\phi)$.
  In addition to possible applications in AdS/CMT, the extremal black branes are of interest from the point of view
of the attractor mechanism. The near horizon  geometries for these branes are universal, independent of the asymptotic
values of the moduli,  and  describe  generic classes of endpoints for
  attractor flows which are different from $AdS_2\times R^2$.}

\preprint{NSF-KITP-09-174, TIFR/TH/09-41,WITS-CTP-046}

\newcommand{\half}{\frac{1}{2}}

\def\be{\begin{equation}}

\def\ee{\end{equation}}

\def\bea{\begin{eqnarray}}

\def\eea{\end{eqnarray}}

\begin{document}

\tableofcontents

\section{Introduction}

\subsection{Extremal dilaton black holes in string theory}

Extremal black holes have been a fruitful source of theoretical questions and enigmas for
several decades.  In the context of black hole mechanics, for instance, they provide examples
where the thermodynamic description breaks down and one is forced to think about the
microphysics of the system in order to make sense of the extremal limit \cite{PSSTW, HW}.

The earliest charged black hole solutions to low-energy string theory were found by 
Garfinkle, Horowitz and Strominger \cite{GHS} (and had appeared earlier in families
of solutions constructed in \cite{GibbonsMaeda}).
Those authors studied the Einstein-Maxwell action with the gauge coupling controlled by a scalar
dilaton $\phi$:
\begin{equation}
\label{GHSaction}
S = \int d^4x ~\sqrt{-g} \left( -R + 2 (\nabla \phi)^2 + e^{-2\phi} F^2 \right)~.
\end{equation}
This action admits remarkably simple extremal magnetically charged black hole
solutions:
\begin{equation}
\label{magmetric}
ds^2 = - (1- {2M \over r})~ dt^2 + {(1 - {2M\over r})^{-1}} ~dr^2 +
r (r- {Q^2 e^{2\phi_0} \over M}) ~d\Omega^2~,~
\end{equation}
with dilaton profile
\begin{equation}
\label{dilaton}
e^{-2\phi} = e^{-2\phi_0} - {Q^2 \over Mr}
\end{equation}
and with gauge field
\begin{equation}
\label{gf}
F = Q ~{\rm sin}(\theta)~ d\theta \wedge d\varphi~.
\end{equation}
Here, $\theta$ and $\varphi$ are standard angular coordinates on the $S^2$ spatial slices
in (\ref{magmetric}), and $\phi_0$ is the asymptotic value of the dilaton.  The action and
the black hole solution are
motivated by the low-energy $\alpha^\prime$ expansion of heterotic string theory; one may obtain an 
equally simple electrically charged solution by exchanging $\phi \to -\phi$, while simultaneously exchanging
the field strength $F$ with its dual $\tilde F_{\mu\nu} = {1\over 2} e^{-2\phi} \epsilon_{\mu\nu}^{~~\lambda\rho}F_{\lambda \rho}$.

The thermodynamics of these black holes, with general dilatonic coupling $e^{-2\alpha\phi} F^2$ 
for $\alpha \geq 0$, was discussed extensively in \cite{PSSTW,HW}.  The black hole solutions in
these theories, for arbitrary values of $\alpha \neq 0$, all share the striking property that the 
entropy at extremality vanishes.  This leads one to suspect that the thermodynamic description
may be breaking down;  and this breakdown can be made precise, arising from two different
sources for $\alpha <1$ and $\alpha \geq 1$.  The basic point is that the number of thermally accessible states of the extremal hole is too small to justify a statistical description, but the root
cause is different in the two cases \cite{HW}:

\noindent
$\bullet$ For $\alpha <1$, these flat-space black holes have positive specific heat and
vanishing temperature as they approach extremality.  (In all cases but $\alpha =0$, the
extremal Reissner-Nordstr\"om case, they also have vanishing entropy).  Thermodynamics
breaks down because there are very few excitations in the relevant thermal interval as
$T \to 0$.  The famous zero temperature successes of statistical physics happen in the
infinite volume limit, while these black holes are fixed objects of a certain size.

\noindent
$\bullet$ For $\alpha \geq 1$, the temperature remains finite (for $\alpha=1$) or even
diverges ($\alpha>1$) as extremality is approached.  However,
${\it the ~black ~hole~develops~a ~mass ~gap}$ which prevents the absorption of
sufficiently small amounts of energy (in units of $k_{B}T$) to justify a thermal
description.  This is evident in the calculation of grey-body factors for the hole, or
equivalently, in the effective Schr\"odinger-like equation which governs absorption and
reflection of radiation by the hole.

In summary, the physics of extremal charged dilaton black holes (like that of extremal Reissner-Nordstr\"om black holes) exhibits a breakdown of the thermodynamic description.  However,
the conceptual reasons for the breakdown depend on details; the relevant physics can be
quite different for different values of $\alpha$.

 Let us end this subsection with one more comment. The  dilaton $e^{2\phi}$ blows up at the horizon of  the extremal magnetically charged black hole
($M=\sqrt{2}Q e^\phi_0$), 
eq.(\ref{magmetric}), eq.(\ref{dilaton}),  which  lies at 
\begin{equation}
r={Q^2 e^{2\phi_0} \over M}.
\end{equation}
Therefore quantum loop corrections will become important close to the horizon. In the electrically charged case in contrast the dilaton vanishes.
However, now the string frame curvature blows up at the horizon and therefore higher derivative corrections 
 will get important close to the horizon. 
This feature is quite common in dilatonic black holes/branes. Near the horizon either the string loop corrections or  the higher derivative  corrections 
 usually become significant and these corrections  bedevil any attempt to calculate properties which depend sensitively on the geometry close to the 
horizon. One way to tame this problem is to consider a slightly non-extremal black hole. For fixed charge and $\phi_0$ the non-extremal hole 
has a horizon at a slightly larger value of $r$ and as a result the dilaton does not run to either infinity or zero value at its horizon. 
 Starting with a large 
enough  charge one needs a temperature which is much smaller than the charge to control the behavior of the dilaton in this way. 
 The properties of the resulting  near-extremal black hole can then be reliably calculated in the classical supergravity approximation.


\subsection{Why pursue their AdS generalization?}

It is of interest to generalize our knowledge of these charged black holes in theories with $\alpha \neq 0$ to similar holes with AdS
asymptotics for several reasons.  Perhaps the most pressing is the recent realization that
AdS/CFT may provide a powerful tool for studying strongly-coupled toy models of
condensed matter systems (for excellent reviews with rather different flavors see \cite{Sachdev,Hartnoll,Herzog,McGreevy}).   At this stage in the development of this
AdS/CMT correspondence, it seems useful to widen the range of qualitative behaviors
seen in simple and potentially relevant gravity models.  In flat space, the extremal dilaton black holes
exhibit several features quite different from their extremal Reissner-Nordstr\"om cousins.
This suggests that their AdS generalizations
should be relevant to capturing holographic phases of matter which are distinct from
those yet explored.

Of particular interest to us at the start of this investigation was the fact that, unlike the
extremal Reissner-Nordstr\"om solutions, these solutions may have vanishing entropy at
extremality.  The large ground state degeneracy of the Reissner-Nordstr\"om AdS black
branes is in tension with the third law of thermodynamics, and suggests to some that
they may be highly atypical holographic states of matter \footnote{
S. Kachru acknowledges several interesting discussions about this  at the June 2009 KITP workshop on ``Quantum Criticality and
the AdS/CFT Correspondence."}$^,$\footnote{The large degeneracy is 
almost certainly an artifact of the 
large $N$ approximation.  SPT acknowledges discussions with A. Dabholkar and A. Sen in the course of preparing \cite{DST},
and also with T. Senthil, in this regard.}. 
Since abelian gauge fields with dilatonic couplings parameterized by various values of $\alpha$ are fairly common
in string compactifications, these may provide one generic class of simple bulk theories where the
charged black holes do not have an undesired macroscopic ground-state entropy.\footnote{In the context of holographic
superconductors \cite{Gubser,H}, another class of black branes with vanishing entropy at $T=0$ was
recently found in \cite{Gauntlett,Gubsertwo,HorowitzRoberts}.}

Another motivation is that the phenomena discussed in \cite{PSSTW,HW} for flat-space
charged dilaton black holes strongly suggest that, at least for some values of the
parameters, charged dilaton black branes in AdS may provide novel holographic
duals of insulators.  While the bulk theory clearly has excitations at arbitrarily low-energy (after all, it has uncharged Schwarzschild black brane solutions), in the sector with non-trivial charge density,
there may be a gap to charged excitations in analogy with \cite{PSSTW,HW}.   We will find that 
this is not so, at least in the absence of a dilaton potential and for the well-motivated simple forms of the gauge
coupling function that we consider.  We will, however, find some other surprises.

A final  motivation for this investigation comes from the study of extremal black hole/branes solutions. 
It is well known by now that these extremal  configurations exhibit the attractor mechanism  
 regardless of 
supersymmetry; their near-horizon geometry is 
universal and independent of the asymptotic values taken by the moduli. Different kinds of attractors correspond
to different kinds of universal  behavior.  
 In the context of AdS/CFT
characterizing the different kinds of attractors  tells us about the  
 different  kinds of IR behavior which can arise in the dual CFT which is at zero temperature but  is  now 
deformed by the addition of a chemical potential (or charge). This is clearly of interest as we develop the 
AdS/CFT dictionary further. From the point of view of  possible connections to condensed matter physics, an  early 
and important paper on the subject, \cite{SachdevSon}, noted  
  that the optical conductivity of the dual CFT in $2+1$ dimensions  at finite temperature (and zero chemical potential) is actually 
independent of  frequency and temperature and thus very universal. 
The attractor mechanism tells us that there should be some considerable universality as we
 deform the CFT along the chemical potential direction instead of temperature as well. And the extent of allowed
variation  should be determined by  the different classes of attractors. Understanding the different classes of attractors
is therefore of interest from this point of view too. A comment on the literature is worth making here. 
There is by now considerable  literature on the attractor mechanism. The seminal paper is \cite{FKS}.
For a recent review  with a good collection of references, see \cite{lectures}. Some references on attractors without 
supersymmetry are \cite{Ga, Gibbons,Sen,attractors,Renata}.

In this paper, we will  study extremal and non-extremal black branes in dilatonic gravity.
In \S2, we find the form of the near-horizon
geometry for electrically charged extremal dilaton black branes as a function of $\alpha$.  
We use this geometry to compute the entropy and specific heat as a function of $\alpha$, and see that the extremal branes have vanishing
entropy and positive specific heat for all $\alpha \geq 0$.   We also discuss other thermodynamic quantities.  
In \S3, we
compute the conductivity in a controlled approximation as one
approaches the extremal limit, using techniques similar to those in \cite{HorowitzRoberts}.  
We find that somewhat surprisingly, in the simple cases we check (including the electrically charged 
black branes for all values of
$\alpha$), the result is that $\sigma(\omega) \sim \omega^2$ at $T=0$ and low frequency.
We conclude with a discussion of the interpretation of our results, and
 of promising directions for future work, in \S4.
In the appendix, we show that one can numerically extend our near-horizon solutions
of \S2\ to provide full black brane solutions with AdS asymptotics.

While this paper was being readied for submission, the papers \cite{Igor,Gubser22}, which have some overlap in motivations with our work, appeared.

\section{Near-horizon behavior of extremal electrically charged brane}

\subsection{Near-horizon solution}

While the full black-brane solutions with AdS asymptotics did not admit a simple analytical form
that we could find, we are able to provide an analytical description of the near-horizon
geometry for the extremal charged dilaton branes \footnote{By the near-horizon region we mean the region 
of spacetime  ``close to'' where the $g_{tt}$ component of the metric   vanishes. }.  We use a notation and formalism similar
to that in
\cite{attractors}, in describing these near-horizon solutions.

For our bulk action, we take
\begin{equation}
\label{ouraction}
S = \int d^4x \sqrt{-g} \left(R - 2(\nabla \phi)^2 - e^{2\alpha\phi} F^2 - 2\Lambda\right)~.
\end{equation}
The maximally symmetric vacuum solution is then of course AdS space with AdS scale
$L$ determined by $\Lambda = -{3 \over L^2}$.
We consider a metric of the form
\begin{equation}
\label{ansatz}
ds^2 = -a(r)^2 ~dt^2 + a(r)^{-2} ~dr^2 + b(r)^2 ~(dx^2 + dy^2)
\end{equation}
and a gauge field of the form
\begin{equation}
\label{gaugefield}
e^{2\alpha \phi} F = {Q\over b(r)^2} ~ dt \wedge dr.
\end{equation}
That is, we are looking for electrically charged black branes.\footnote{Related work finding charged black brane solutions in dilaton gravity with a Liouville-type potential
appears in e.g. \cite{dbh}.}

The equations of motion can be easily inferred from (107)-(110) of \cite{attractors}.  We find
that
\begin{equation}
\label{first}
(a^2 b^2)'' = - 4\Lambda b^2
\end{equation}

\begin{equation}
\label{second}
{b'' \over b} = - (\partial_r \phi)^2
\end{equation}
\begin{equation}
\label{third}
\partial_r (a^2 b^2 \partial_r \phi) = -{\alpha}e^{-2\alpha\phi} {Q^2 \over b^2}~,
\end{equation}
together with the first order constraint,
\begin{equation}
  \label{constraint}
  a^{2}b'^{2}+\half {a^{2}}'{b^{2}}'=\phi'^{2}a^{2}b^{2}-e^{-2\alpha\phi}\frac{Q^{2}}{b^{2}}-b^{2}\Lambda.
\end{equation}

Even though we will be interested in the near-horizon limit, it is important to keep the
cosmological constant term on the right-hand side of (\ref{first}); the black brane
(unlike the black hole) has vanishing curvature along the $x,y$ spatial slices, so the
cosmological constant does provide the most significant source in some of the Einstein
equations. As was mentioned above, 
the radius of $AdS$ space is given by
\begin{equation}
\label{defradius}
L=\sqrt{-3/\Lambda}~.
\end{equation}
In the discussion below unless  specified otherwise we will  set $L=1$. When needed the  dependence on $L$ can  be determined by dimensional analysis. 

To find the near-horizon limit, we proceed with a scaling ansatz.  Define the near-horizon
variable $w = r - r_H$ where $r_H$ is the radius of the horizon.  Let us make the ansatz
that the near-horizon scalings of the functions $a, b, \phi$ are given by
\begin{equation}
\label{scaling}
a =C_2 w^{\gamma}, ~b =C_1 w^{\beta}, ~\phi = -~K~ {\rm log}(w)+C_3~,
\end{equation}
where $C_1,C_2,C_3$ are constants. 
A little algebra then shows that an {\it exact} solution to the 
 equations (\ref{first}), (\ref{second}), (\ref{third}) and the constraint eq.(\ref{constraint})
 is obtained if the exponents  take the values
\begin{equation}
\label{exponents}
\gamma = 1, ~K = {{\alpha \over 2} \over {1 + {({\alpha \over 2})}^2}},~
\beta = {({\alpha \over 2})^2 \over {1 + ({\alpha \over 2})^2}}~.
\end{equation}

The constant $C_2$ is determined to be
\begin{equation}
C_2^2 =  {6\over(\beta+1)(2\beta+1)}~.
\end{equation}
By rescaling $w$, $t$, $x$ and $y$ appropriately the constant $C_3$ can be set to zero and $C_1$ to unity. 
One then gets that  $Q$ is determined in terms of $\alpha$ by
\begin{equation}
\label{constants}
Q^2  =  { 6 \over (\alpha^2+2)}
\end{equation}

The final values of the metric components are then
\begin{equation}
\label{attractora}
a =C_2 w, ~b = w^{\beta}, ~\phi = -~K~ {\rm log}(w)~.
\end{equation}

Note that since $a$ vanishes linearly with $w$, the metric component $g_{tt}$ has a second order zero at $w=0$, as is needed for an extremal solution. 
In the extremal solution found above  the gauge coupling $g_{\rm U(1)}\sim e^{-\alpha\phi}$ becomes arbitrarily weak at the horizon,
as $\phi \to \infty$. Conversely, when $w\rightarrow \infty$ the gauge coupling becomes very  strong.

The scaling exponents  that characterize the solution actually indicate that in the near-horizon region, the metric takes the
form characterizing a Lifshitz fixed point with anisotropic scaling \cite{KLM} (see also \cite{other}).  The dynamical
critical exponent is fixed to $z = {( 1 + ({\alpha \over 2})^2 ) \over ({\alpha \over 2})^2} = 1/\beta$.
The scaling symmetry is not exact, it is broken by the  logarithmic dependence of the 
dilaton on $w$. 

As was mentioned above the scaling solution eq.(\ref{scaling}), eq.(\ref{exponents}), eq.(\ref{constants}), is an exact solution to the equations of motion. 
However for our purposes it does not have the correct  asymptotic behavior.\footnote{Solutions of similar systems with this kind of asymptotic behavior
will appear in \cite{johnnew}, where they can be interpreted in terms of a microscopic dual theory with Schr\"odinger symmetry.}
We are interested in solutions which asymptote to $AdS_4$. 
In the \S2.4 and in appendix A, we will discuss how a new solution can be obtained after adding a perturbation to the above solution.
This solution  is  asymptotically $AdS_4$  and has  an asymptotically  constant dilaton. After a coordinate transformation we see that the 
charge $Q$ is fixed to a universal value in the scaling solution,  eq. (\ref{constants}). In contrast in the asymptotically $AdS_4$
case the charge $Q$ will be directly related to the number density in the dual theory \footnote{This follows directly from the gauge field
eq. (\ref{gaugefield}) and the $AdS/CFT$ dictionary.}, and has physical significance.
In addition the $AdS_4$ solution will also depend on $\phi_0$ -- the asymptotic value of the dilaton. 
The asymptotically $AdS_4$ solution  will therefore be characterized by two parameters, $Q$ and $\phi_0$.

The solution we have discussed above,  with a Lifshitz-symmetric metric,  was in fact already found in \S3\ of \cite{Taylor}, where
however the focus was on black brane solutions with different (Lifshitz instead of
$AdS$) asymptotics \footnote{Other
interesting
examples of black hole solutions which asymptote to 4D Lifshitz metrics with various
values of the dynamical exponent $z$ have appeared in
\cite{LifBH}.}.

\subsection{Attractor behavior}

Note that the solution in eq.(\ref{attractora}) has no free parameters. The exponents $\beta, K,$ and the constant $C_2$ which appear in it are  
determined in terms of the parameter $\alpha$, as in eq. (\ref{exponents}) and eq. (\ref{constants}). We remind the reader that $\alpha$ 
 appears in the Lagrangian, eq. (\ref{ouraction}), and  
 determines the gauge coupling in terms of the dilaton.  Similarly the parameter $Q$ which appears in the gauge field eq. (\ref{gaugefield})
 is not a free parameter in the scaling solution and  is also 
    determined in terms of $\alpha$, eq. (\ref{constants}).  In contrast the full solution which is asymptotically $AdS_4$ has 
two free parameters which are needed to specify it, 
the charge $Q$ and the asymptotic value of the dilaton, 
$\phi_0$, as was mentioned above
and as we will see in greater detail in section 2.4. 
 We see that regardless of the values 
both of these parameters take in the full solution, the near-horizon region of the solution takes a universal form. 
This  universality is a result of the attractor mechanism,  although 
now at work in a 
somewhat less familiar situation where the attractor geometry does not have an $AdS_2$ factor giving rise to an  $SO(2,1)$ isometry.

Let us elaborate   on this connection with the attractor mechanism further. 
The equations of motion, eq.(\ref{first}), eq.(\ref{second}), eq.(\ref{third}), can be obtained from a one dimensional action, 
\begin{equation}
\label{action}
S=\int dr \big(-2a^2b b''-2a^2b^2(\partial_r\phi)^2 -2{V_{eff}\over b^2}+{6b^2\over L^2}\bigr)
\end{equation}
with  the effective potential $V_{eff}$ being 
\begin{equation}
\label{effectivepot}
V_{eff}=e^{-2\alpha \phi} Q^2.
\end{equation}
In addition the constraint eq.(\ref{constraint}) must also be satisfied.  In terms of $V_{eff}$ this is the condition,
\begin{equation}
  \label{constraintgen}
  a^{2}b'^{2}+\half {a^{2}}'{b^{2}}'=(\phi_i')^{2}a^{2}b^{2}-\frac{V_{eff}}{b^{2}} +3 {b^{2}\over L^2}.
\end{equation}
 
Consider a more general situation now with  a four dimensional Lagrangian of the form,
\begin{equation}
\label{gen4dimact}
S=\int d^4x\sqrt{-g}\bigl(R-2\Lambda-2(\partial\phi_i)^2 -f_{ab}F^aF^b-{1\over 2} \tilde{f}_{ab}\epsilon_{\mu\nu\rho\sigma} F^a_{\mu\nu}F^b_{\rho\sigma}
\bigr )
\end{equation}
which has $i=1 \cdots N$  scalars  with standard kinetic energy terms and $a=1, \cdots M$ gauge fields.\footnote{Incorporating
 a  more general kinetic energy term for the scalars, as would arise  from a general K\"ahler potential in a supersymmetric theory,
 is straightforward but  we do not discuss it further here.}
And take a case where both the  electric and magnetically charges are excited.  One finds in a coordinate system of the form, 
eq.(\ref{ansatz}) that the gauge fields are  given by 
\begin{equation}
\label{gengaugef}
F^a=f^{ab}(Q_{eb}-\tilde{f}_{bc}Q_m^c){1\over b^2}dt\wedge dr+Q_m^a dx\wedge dy
\end{equation}
with $Q^a_m, Q_{ea}$ being constants that determine the electric and magnetic charges of the system and $f^{ab}$ being the inverse of the 
gauge coupling function $f_{ab}$. The  equations of motion for the metric and scalars can be obtained by varying the action eq.(\ref{action})
 with the effective potential now being given by 
\begin{equation}
\label{genefpot}
V_{eff}=f^{ab}(Q_{ea}-\tilde{f}_{ac}Q^c_m)(Q_{eb}-\tilde{f}_{bd}Q_m^d)+f_{ab}Q_m^aQ_m^b.
\end{equation}
Similarly the constraint takes the same form eq.(\ref{constraintgen}) with this new effective potential.  

In the standard attractor situation, $V_{eff}$ has a critical point at some finite point in moduli space. The resulting extremal black brane 
has a horizon where the scalars are drawn to their critical values, $\phi_{i*}$, regardless of their asymptotic values at infinity. The metric component
$a^2$ has a second order zero at the horizon, while the metric component $b^2$ attains a non-zero value $b_h^2$ at the horizon. 
As a result the near-horizon geometry is of the form $AdS_2\times R^2$ and has an $SO(2,1)$ isometry which arises from the $AdS_2$ factor. 
The entropy density of the 
black brane $s\sim   b_h^2$. From the constraint eq.(\ref{constraintgen}) it follows that $b_h^2$   is determined 
by the effective potential at the critical point,
\begin{equation}
\label{valbhor}
b_{h}^4={L^2 V_{eff}(\phi_{i*}) \over 3}~.
\end{equation}

In contrast in the  situation we have encountered above eq.(\ref{ouraction})
 the effective potential is of ``run-away" form, with a critical point
 which lies at infinity. Also  the   critical value of the effective potential vanishes, and this value is obtained exponentially rapidly in the dilaton. As a result of these properties
 the near horizon geometry is of the scaling type found 
above, where the scalar runs towards but never quite gets  to its   critical value  
and where the entropy vanishes. It is easy to see that from the equation of motion, eq.(\ref{first})-eq.(\ref{constraint})
that  the full solution for the metric
 (which is asymptotically $AdS_4$)
only depends on the combination $Q^2e^{-2\alpha \phi_0}$, where $\phi_0$ is the asymptotic value of the dilaton. 
But the attractor mechanism implies that the near-horizon metric is independent of the dilaton $\phi_0$.   It then also imples that the scaling solution 
is independent of
 the charge parameter $Q$,  which is consistent with our previous explicit result eq.(\ref{attractora}).

The scaling solution we have obtained can  arise as an  attractor in 
other situations as well.\footnote{Actually here we will only show that the  same scaling solution can 
arise  in other systems  as well. Whether this happens starting in the near-horizon region 
of a solution with say $AdS_4$ asymptotics is a more involved question that the comments below 
will not address. This is also true about our comments below when we discuss other kinds of  attractors.}
For example, consider a situation with several scalars, and a run-away potential which depends on a linear combination of these scalars,
\begin{equation}
\label{lincomb}
V_{eff}=V_0 e^{-\alpha_i\phi_i ~.}
\end{equation}
After a field redefinition this maps to the single scalar case above. 
As another  example consider an  effective potential which has  a critical point 
  at a finite value in field space for all but one of the scalars. 
The one remaining scalar runs away to infinity driving the potential to zero in an exponentially rapid fashion. 
In this case the same attractor geometry, eq.(\ref{scaling}),  with appropriate scaling exponents, will arise. 
Finally consider  a situation  where there are several scalars with a  potential
\begin{equation}
\label{genefra}
V_{eff}=\sum_iV_i e^{-2\alpha_i\phi_i}
\end{equation}
with all $V_i>0$.
Again the near horizon region  takes the same form as in eq.(\ref{scaling}) 
\begin{equation}
\label{srattractor}
a\sim w, b\sim w^\beta, \phi_i\sim -K_i \log(w) + C_i~,
\end{equation}
now with:
\begin{eqnarray}
\label{genexponents}
\beta &=&{1\over (1+4\sum{1\over \alpha_i^2})} \\
K_i &= & {2\beta\over \alpha_i} ~.
\end{eqnarray}

\subsubsection{Other kinds of attractors}

To complete this discussion let us also consider other possible attractor solutions which can arise. 
We list some of these possibilities below. 
We work in the coordinate system eq.(\ref{ansatz}) below with $w=r-r_h$. 
In all the cases we consider below  $a \sim w$ in the near horizon region, so that the 
$g_{tt}$ component has a second order zero at $w=0$. 
As was mentioned above, the scaling solution eq.(\ref{scaling}) is an exact solution to the equations of motion. 
In contrast the solutions we write down below will not in general be exact, rather we will give the leading behaviour
in the near horizon region for the metric and dilaton in these cases.

1) Suppose the effective potential takes the form,
\begin{equation}
\label{formep3}
V_{eff}=V_0+V_1e^{-2\alpha\phi}
\end{equation}
where $V_0, V_1>0$. 
This results in a run-away situation, but the critical value of $V$ is now $V_0$ and does not vanish. 
In this case one finds  that 
\begin{eqnarray}
\label{exp3}
b & = & b_h+{C_1\over \log(w)} \\
 b_h^4 & = &  {L^2 V_0 \over 3} \\
\phi & = & {1\over 2\alpha}\log(-\log(w)). 
\end{eqnarray}
Since $a\sim w$ and $b_h \ne 0$ the near horizon geometry is $AdS_2\times R^2$.  Interestingly though
 the scalar does not become a constant 
in the near-horizon region going instead  to infinity
as $w\rightarrow 0$. 

2) Next consider the case where the potential vanishes at the critical point but as a power-law  rather than 
 an exponential:
\begin{equation}
\label{efpot4}
V_{eff}={V_0 \over  \phi^p}.
\end{equation}
Now one finds  
\begin{eqnarray}
\label{exp4}
b & \sim &  {1\over \log(w)^{p\over 8}} \\
\phi& \sim & (-\log(w)) ^{1/2}~.
\end{eqnarray}
The power law nature of the potential results in $b$ and $\phi$ varying more slowly than 
in the scaling solution eq.(\ref{scaling}).

3) We can also  contrast  this with a  potential  that vanishes more rapidly than an  exponential:
\begin{equation}
\label{efpot5}
V=Q^2 e^{-A(e^{\alpha \phi})}.
\end{equation}
One finds now that 
\begin{eqnarray}
\label{exp5}
b  & \sim & w+{C_1 w \over \log(w)} \\
\phi & \sim & {1\over \alpha} \log \left(-\frac{1}{A}\log \left( h(w) \right) \right) \\
h(w) & = & \frac{3}{2Q^2\alpha^2} w^4(-\log (w))^{-2} + \ldots 
\end{eqnarray}

The metric component $b$ is almost linear  in $w$ resulting in the near horizon geometry being approximately $AdS_4$,
with a very slowly varying scalar. 

4) Finally one can consider a situation where the  four-dimensional theory one starts with has a potential which depends on the scalar field.
We write the one dimensional Lagrangian as
\begin{equation}
\label{newone}
S=\int dr \left(-2a^2bb''-2a^2b^2(\partial \phi)^2-2{V_{eff}\over b^2}+ 6 {b^2\over L^2} -2 V_1(\phi) b^2 \right)
\end{equation}
where $V_1(\phi)$ is the extra field-dependent potential. 
There are now several possibilities. Let us only discuss one of these here. 

An $AdS_2\times R^2$ solution arises, where the metric component $b^2$, eq.(\ref{ansatz}),  takes a constant value $b_h^2$ 
and the scalar takes a constant value $\phi_*$,  if $b_h$ and $\phi_*$ can be found which solve the two equations, 
\begin{equation}
\label{conda1}
{\partial_\phi V_{eff}(\phi_*) \over b_h^2} +\partial_\phi V_1(\phi_*) b_h^2=0
\end{equation}
and
\begin{equation}
\label{condb}
({3 \over L^2}-V_1(\phi_*))b_h^4=V_{eff}(\phi_*).
\end{equation}

\subsection{Thermodynamics of the near-extremal solution}

A generalization of the scaling solution can be found for the action eq.(\ref{ouraction}) \cite{Taylor}. It has a  metric
\begin{equation}
\label{genscaling}
ds^2=C_2^2 w^2(1-{m \over w^{2\beta+1}}) dt^2+     
{dw^2 \over C_2^2 w^2(1-{m \over w^{2\beta+1}})}   + w^{2\beta}(dx^2+dy^2),
\end{equation}
which now depends on the parameter $m$. The constant $C_2$ takes the same value as in eq.(\ref{constants}). 
 The dilaton and gauge field are unchanged from their values in the scaling solution and are given in 
eq.(\ref{attractora}), eq.(\ref{gaugefield}),
eq.(\ref{constants}), respectively.   
Asymptotically, as $w\rightarrow \infty$,   this  solution reduces to the original  scaling solution eq.(\ref{attractora}). 
The behavior close to the horizon though is different. The $g_{tt}$ component of the metric now has a first order zero, with non-vanishing surface gravity,
and as a result  the resulting temperature in non-zero. The scaling solution in eq.(\ref{attractora})
 corresponds to the near-horizon of an extremal black brane. 
We therefore expect that the  new solution above  corresponds to the near-horizon region of a slightly non-extremal black brane. 

The horizon in the solution eq.(\ref{genscaling}) is located at $w_h$, where $w_h$ satisfies
\begin{equation}
\label{genhor}
w_h^{2\beta+1}=m.
\end{equation}
The resulting temperature which can be obtained in the standard fashion by continuing to Euclidean space \cite{GibbonsH} is
\begin{equation}
\label{temp}
T\sim w_h.
\end{equation}
And the entropy density  is then
\begin{equation}
\label{entropyh}
s\sim w_h^{2\beta}\sim T^{2\beta}. 
\end{equation}

As was mentioned above the solution eq.(\ref{genscaling}) arises as the near horizon limit of a slightly non-extremal black brane solution. 
The  entropy density  can be expressed as a function of two dimensionful parameters for the non-extremal solution,
 the  temperature $T$ and the chemical potential $\mu$.
Both of these are intensive variables with dimensions $[Mass]^1$. 
It follows from eq.(\ref{entropyh}) and dimensional analysis  that the entropy density of a slightly non-extremal black brane is given by
\begin{equation}
\label{entropysne}
s\sim T^{2\beta} \mu^{2-2\beta}~.
\end{equation}

The entropy can be found from the classical action eq.(\ref{action}) evaluated on-shell. This action can be calculated by scaling out the dependence 
on $L$ - the radius of AdS space -  and then working with dimensionless quantities. As a result the action and the entropy will have a prefactor 
which goes 
like,  $L^2/G_N$ where $G_N$ is the four dimensional Newton's constant. 
Putting all this together gives the entropy density of the slightly non-extremal black brane to be
\begin{equation}
\label{entropyfinal}
s= a C T^{2\beta} \mu^{2-2\beta},
\end{equation}
where  
\begin{equation}
\label{valc}
C \sim L^2/G_N
\end{equation}
is the central charge of the CFT dual to the $AdS_4$ background. 
The coefficient $a$ depends on $\alpha$, and $\phi_0$, 
 this dependence can be fixed from the numerical solution for the slightly non-extremal black hole.

Note that the specific heat  
\begin{equation}
\label{specificheat}
C_v=T ({ds \over dT})_\mu= (2\beta) a C   T^{2\beta} \mu^{2-2\beta}
\end{equation}
is  positive. 
 
The other  thermodynamical properties  can be completely determined from the entropy density. 
The Gibbs-Duhem relation
\begin{equation}
\label{gibbsduhem}
sdT-dP + n d\mu=0,
\end{equation}
where $P, n,$ are the pressure and number density, can be used to obtain the pressure. Keeping $\mu$ fixed and integrating the above equation gives,  
\begin{equation}
\label{pressure}
P={a\over (2\beta+1)}C \mu^{2-2\beta} T^{2\beta+1} + b C e^{3\alpha \phi_0} \mu^3.
\end{equation}
The second term is a temperature independent integration constant, in \S2.4 we will see that the coefficient $b$ is indeed non-zero 
and determine its scaling with $\phi_0$. 
Substituting for $P$ in eq.(\ref{gibbsduhem}) gives the number density, 
\begin{equation}
\label{numberdensity}
n={(2-2\beta) \over (2\beta+1)} a C T^{2\beta+1} \mu^{1-2\beta} + 3 b C e^{3\alpha \phi_0} \mu^2. 
\end{equation}

Finally the energy density can be obtained using the relation
\begin{equation}
\label{rel2}
\rho=sT+\mu n-p
\end{equation}
which gives,
\begin{equation}
\label{mass}
\rho= {2 a\over (2\beta+1)}C \mu^{2-2\beta} T^{2\beta+1} + 2b C e^{3\alpha \phi_0} \mu^3 .
\end{equation}
We see from eq.(\ref{pressure}) that the near-extremal system has an equation of state
\begin{equation}
\label{estate}
P={1\over 2} \rho~.
\end{equation}

The formulae obtained in this section are  valid when $ T \ll \mu $, i.e. for a slightly non-extremal black brane. 
As the temperature increases for fixed $\mu$ the geometry eq.(\ref{genscaling})
 is no longer a good approximation and the corrections to the formulae
above become significant. 

Let us end this section with some additional  comments.
 We have seen that the specific heat, which governs the fluctuations in energy, is positive. 
From eq.(\ref{numberdensity}) we can also compute the susceptibility $\chi$. 
One gets,  
\begin{equation}
\label{susep}
\chi\equiv ({\partial n \over \partial \mu })_T = (1-2\beta){(2-2\beta) \over (2\beta+1)} a C T^{2\beta+1} \mu^{1-2\beta} 
+ 6 b C \mu 
 \end{equation}
For $T \ll \mu$ this is always positive. It is worth noting that the first term, which is growing with 
temperature, is negative for $\beta > 1/2$. Whether the susceptibility actually turns negative though as the temperature increases,
signalling a that phase transition occurs, 
 requires one to go beyond the regime where $T \ll \mu$. We leave this question for the future \footnote{We thank K. Damle and S. Minwalla
for related discussions.}. 

The scaling, for example of the entropy density with temperature, follows essentially from the Lifshitz-like nature of the 
near-horizon region. The scale invariance of the metric in  this region is broken only by the temperature and this together with dimensional analysis then 
determines the temperature dependence in eq.(\ref{entropyh}). 
In turn then the temperature dependence of other thermodynamic variables also follows from the dynamic exponent being $1/\beta$ in the scaling region. It is also interesting to note that  for $\beta = {1\over 2}$, the thermodynamics 
of this system is quite similar to that of a free Fermi gas.\footnote{We thank K. Rajagopal
for making a related helpful comment during a seminar about this work at MIT.} However, we expect that correlations 
functions, for example the two point density-density correlation function,  can distinguish between these two possibilities
and will agree with that of a Lifshitz-like theory \footnote{We thank K. Damle for suggesting this. 
It is also possible that the ratios of coefficients in various thermodynamical parameters will be different. 
We have not carefully computed them here.}.

The properties we have found above, notably the vanishing entropy and temperature and positive specific heat for the charged dilaton black branes,  
are to be contrasted with
the properties of various other charged black holes.   For instance, AdS Reissner-Nordstr\"om
black holes (as with Reissner-Nordstr\"om black holes in flat space) have a large entropy at extremality; the extremal string-theoretic holes of \cite{GHS} have
finite temperature and negative specific heat; and the more general flat-space black holes with arbitrary $\alpha$ have
negative specific heat for $\alpha \geq 1$.  
The black branes we have found seem most similar, then, to the $0 < \alpha < 1$ flat-space
dilaton black holes studied in \cite{PSSTW,HW}.  They have positive specific heat and vanishing
entropy in the extremal limit. 

It is of interest now to perturb the near-horizon geometry with a background electric field, and ask about charge transport.  At vanishing charge density, the
black brane solution is the standard AdS-Schwarzschild solution, and the conductivity is a constant at small $\omega$ and $T=0$.   (The transport in such
backgrounds has been explored in detail in \cite{SachdevSon}, where in fact the authors argue that $\sigma(\omega/T)$ is a constant function).
We will find that in contrast, a background charge density gives rise to a universal, $\alpha$-independent
behavior $\sigma(\omega) \sim \omega^2$ at $T=0$; the $T$-dependent DC conductivity also has non-trivial corrections to the constant form
found in \cite{SachdevSon}, though we shall not detail them here.  Before performing transport calculations, however, we finish constructing the global charged black brane solutions.

\subsection{Obtaining the solution with the correct asymptotics}

The scaling solution eq.(\ref{attractora}) does not have the correct asymptotic behavior as we have emphasized already.   While the near-horizon behavior sufficed
for us to glean the most interesting thermodynamic behaviors in \S2.3, it is of interest to find the structure of the global solutions.
Here we will describe some of the details which go into finding a solution which does asymptote to $AdS_4$. More details can be found in appendix A. 
The idea is to add a perturbation to the scaling solution which is irrelevant at small $w$, close to the horizon, but which gets increasingly important at 
larger values of $r$. Such a perturbation if correctly chosen can then change the solution giving rise to the required new solution which asymptotes to 
$AdS_4$. Roughly speaking one can think of large $r$ as the ultraviolet and small $r$ as the infrared in the boundary theory. Thus the perturbation we add
is irrelevant in the infrared scaling region but relevant in the UV. 
In the discussion below we will work in conventions where $L=1$. 

The perturbation we consider preserves the form of the metric eq.(\ref{ansatz}) (and thus the symmetries of eq.(\ref{ansatz})). 
The resulting functions $a, b, \phi$ then are:
\begin{eqnarray}
\label{perta}
a& = & C_2 w (1+d_1 w^\nu) \\
b& = & w^\beta(1+d_2 w^\nu) \\
\phi& = & -K\log (w) + d_3 w^\nu 
\end{eqnarray}
$C_2,K,\beta$ take the same values as in eq.(\ref{constants}), eq.(\ref{exponents}). There is a gauge field turned on, eq.(\ref{gaugefield}) 
with $Q$ given by eq.(\ref{constants}). 
The perturbation is characterized by the exponent $\nu$ and the three constants $d_1, d_2, d_3$. 
As we see in appendix A  $d_2, d_3$ can be determined in terms of 
$d_1 \equiv d$. And requiring that $\nu>0$, so that the perturbation dies out at small $w$,  gives rise to a unique allowed value for this exponent:
\begin{equation}
\label{expgamma}
\nu={-(2\beta+1) + \sqrt{(2\beta+1)(10\beta+9)} \over 2}~.
\end{equation}
Thus there is a one parameter family of allowed perturbations, parameterized by the coefficient $d$. 
Adding this parameter with $d<0$ gives rise to an asymptotically $AdS_4$ solution. We verify this by numerically integrating the equations of motion. 

We expect that the boundary theory at zero temperature is characterized by two parameters, the value of the chemical potential (or charge density) 
and the asymptotic value of the dilaton. 
The discussion above though seems to yield only a one parameter family of  solutions, up to coordinate transformations. The answer to this puzzle is 
tied to the fact that the behavior of the  boundary theory for different values of the chemical potential can be obtained by a rescaling, since 
the chemical potential is the only scale in the boundary theory. 
 Now   a scaling transformation in the boundary theory is actually a coordinate transformation in the bulk which involves a rescaling of the radial variable.  
Thus to allow for a change in chemical potential one must treat  bulk  solutions related by a rescaling coordinate transformation as being distinct.
This then introduces a second parameter in the bulk solutions as well. 

In practice there is one bulk solution from which all others can be obtained by a suitable  rescaling and  shift in the dilaton. 
 It is clear from the equations of motion that  the metric and $\phi-\phi_0$  only depend on $Q^2e^{-2\alpha \phi_0}$. 
The rescaling is obtained by carrying out a coordinate transformation under which, 
\begin{equation}
\label{rescale}
r\rightarrow \lambda r, (t,x,y) \rightarrow \lambda^{-1}(t,x,y)
\end{equation}
 Thus starting from the solution for $Q=1, \phi_0=0$ one can rescale to get any value of $Q$ and then shift the dilaton to get any value of $\phi_0$.

It follows from the equations of motion eq.(\ref{first})-eq.(\ref{third}) and the constraint eq.(\ref{constraint})
 that in the  asymptotic $AdS_4$ region the bulk solution, with gauge field, 
eq.(\ref{gaugefield}), the metric and dilaton  must take the form
\begin{eqnarray}
\label{assexp}
a^2& = & r^2(1-e_1 {\rho \over r^3} + {Q^2e^{-2\alpha \phi_0} \over r^4} + \cdots)\\
b^2& =& r^2(1+ \cdots) \\
\phi&=&\phi_0+{\phi_1\over r^3}+ \cdots,
\end{eqnarray}
where the ellipses denote terms that are subdominant at large $r$. 
$\rho$ above is the energy density of the brane.  And $e_1$ is a  constant which depends on $L$.
Under the rescaling eq.(\ref{rescale}), $\rho \rightarrow {\rho \over \lambda^3}, Q\rightarrow {Q\over \lambda^2}$. This tells us that 
\begin{equation}
\label{reeq}
\rho =D_1 (Q e^{-\alpha \phi_0})^{3/2}.  
\end{equation}
The coefficient $D_1$ is a coefficient that is $\alpha$ dependent. 
A similar scaling argument tells us that the chemical potential 
\begin{equation}
\label{chempot}
\mu=\int_{r_h}^\infty {Qe^{-2\alpha \phi}\over b^2} dr
\end{equation}
is given by
\begin{equation}
\label{chempot2}
\mu=D_2 (Q e^{-\alpha \phi_0})^{1/2} e^{-\alpha \phi_0}
\end{equation}
where $D_2$ is again an $\alpha$ dependent coefficient. 
This gives
\begin{equation}
\label{encp}
\rho=D_3 e^{3\alpha \phi_0} \mu^3.
\end{equation}
The coefficient 
\begin{equation}
\label{cd3}
D_3 \propto C
\end{equation}
where $C$ is the central charge  eq.(\ref{valc}) and the proportionality constant which can be extracted from the numerical solution is 
denoted by $2b$ in eq.(\ref{mass}).  

Let us also note that from these relations (and also more directly from the gauge field eq.(\ref{gaugefield}) and the standard
 AdS/CFT dictionary which relates the boundary value of the gauge field to the charge density) it follows that  
  the  number density satisfies the relation
\begin{equation}
\label{numberdensityb}
n\propto Q.
\end{equation}

We end this section with one final comment. 
We had mentioned at the end of section 1.1 that classical relativity often breaks down in the near horizon region of 
extremal dilatonic black holes, either due to loop corrections or higher derivative corrections becoming important. 
The electrically charged extremal branes  we are considering in this paper are the analogue of the electrically charged dilatonic 
black hole of section 1.1. The gauge coupling $g^2=e^{-2\alpha \phi}$ becomes  vanishing small 
 in these cases in the  near horizon region. 
Typically this would lead to higher derivative  corrections becoming important. We had also mentioned in section 1.1 that 
this problem can be dealt with by introducing a small non-zero temperature. From eq.(\ref{scaling}),  eq.(\ref{temp}) it follows that with $T\ne 0$
the dilaton at the horizon is  
\begin{equation}
\label{dilhft}
e^{-2\alpha \phi} \sim \left({T\over \mu}\right)^{4\beta}.
\end{equation}
Thus the dilaton is prevented from running to zero at  non-extremality. In addition  large $\mu$ or charge 
and an adjustable asymptotic value of the dilaton, which we have suppressed in eq.(\ref{dilhft}),
   can also help in obtaining 
a small curvature.  

Below we will compute the conductivity in the extremal case. The near-horizon region will play an important role in this calculation
and one should   worry about corrections that would arise in any genuine string theory embedding of such a black brane.
Once again the safest way to make the calculation reliable is to introduce a small non-zero temperature and then calculate 
the conductivity in the frequency range, $T \ll \omega \ll \mu$. In this case we expect that 
a small nonzero temperature will control
the corrections, and  since the frequency is much larger than the temperature, the conductivity should essentially agree with the calculation
done in the extremal black brane background.

\section{Computing the conductivity at zero-temperature}

\subsection{Finding an effective Schr\"odinger problem}

We now compute the behavior of the conductivity $\sigma(\omega)$ in the extremal black brane background. 
We use the general formula for $\sigma$ recently derived in 
\cite{HorowitzRoberts}, and crucially rely on appropriate generalizations of some of the equations from
\cite{GubserRoca,HHH}.

The conductivity can be computed by turning on a component of the $U(1)$ gauge field
parallel to the black brane, and studying an appropriate two-point function.   A general
discussion with further references appears in e.g. \cite{Kovtun}.  A useful formulation of
the conductivity for our purposes was given in the paper \cite{HorowitzRoberts}.  After
 turning on a gauge field $A_{x}(r,t)$ in the black-brane background, one recasts the second order differential equation
for the perturbation as a Schr\"odinger-like equation:
\begin{equation}
\label{Seq}
-A_{x,zz} + V(z) A_x = \omega^2 A_x
\end{equation}
where $z$ is a redefinition of the radial variable, chosen to cast the equation of motion for
the perturbation in the form (\ref{Seq}), and we have taken a single Fourier component in
frequency space (so $A_x = A_x(z)$).  Then, studying scattering with purely incoming boundary
conditions at the horizon, one can compute the conductivity $\sigma(\omega)$ in terms of
the reflection coefficient

\begin{equation}
\label{condis}
\sigma(\omega) = {{1 - {\cal R}}\over {1 + {\cal R}}}~.
\end{equation}
We will see that in our setup, we can derive analogous equations, though there are some
minor complications, and hence modifications of the formula (\ref{condis}).

It remains to find the equation for perturbations $A_x$ and cast it in the form (\ref{Seq}). 
It will prove simplest to work in the metric gauge used by \cite{HorowitzRoberts}:

\begin{equation}
\label{newgauge}
ds^2 = -g(r) e^{-\chi(r)} ~dt^2 + {dr^2\over g(r)} + r^2 (dx^2 + dy^2)~.
\end{equation}
It is important not to confuse the radial variable $r$ here with the natural radial variable
in \S2; they are quite different.

In this gauge, the equation of motion for the Maxwell field, assuming the
Lagrangian takes the general form 
\begin{equation}
\label{ourac}
{\cal L} = ... - 2 (\nabla\phi)^2 - {1\over 4}f^2(\phi) F_{\mu\nu}F^{\mu\nu} + ...
\end{equation}
where the gauge-coupling function is $f^2(\phi)$, is given by
\begin{equation}
\label{gfeom}
\partial_r \left((f(\phi))^2 e^{-{\chi\over 2}} g \partial_r A_x \right) +
\omega^2 e^{\chi \over 2} g^{-1} A_x (f(\phi))^2 +
(f(\phi))^2 e^{\chi \over 2} (\partial_r A_t) \left( g'_{xt} - {2\over r} g_{xt} \right) = 0~.
\end{equation}
Here we have anticipated, following \cite{GubserRoca,HHH}, that $g_{xt}$ will be turned on at the
same order as the gauge field perturbation; and we are working to lowest nontrivial order
in $A_x$ and $g_{xt}$.

The $xr$ component of the Einstein equations tells us that

\begin{equation}
\label{Exr}
2 {e^{\chi\over 2}\over g} \left( g'_{tx} - {2\over r} g_{tx} + (f(\phi))^2 (\partial_r A_t) A_x
\right) = 0~.
\end{equation}
Using (\ref{Exr}) in (\ref{gfeom}), we find

\begin{equation}
\label{simpler}
\partial_r \left( (f(\phi))^2 e^{-{\chi \over 2}} g \partial_r A_x \right) + \omega^2
e^{\chi \over 2} (f(\phi))^2 g^{-1} A_x - (f(\phi))^4~ e^{\chi \over 2}
(\partial_r A_t)^2 A_x  = 0~.
\end{equation}

Finally, let us define a new variable

\begin{equation}
\label{zis}
{\partial \over \partial z} = e^{-{\chi\over 2}} g {\partial \over \partial r}~,
\end{equation}
and a new wavefunction
\begin{equation}
\label{psiis}
\Psi = f(\phi)~ A_x~.
\end{equation}
It follows from (\ref{simpler}) that $\Psi$ satisfies a Schr\"odinger equation

\begin{equation}
\label{finalSeq}
- \Psi^{\prime\prime} ~+~V(z) \Psi~=~\omega^2 \Psi
\end{equation}
where the potential is given by
\begin{equation}
\label{nowpot}
V(z) = (f(\phi))^{-1} \left( (f(\phi))^{\prime\prime} +  g^{-1} f^3(\phi) e^{\chi} (A_t^\prime)^2 \right)~
\end{equation}
and $^\prime$ denotes ${d\over dz}$.
In other words, we will re-formulate the calculation of the conductivity in terms of the study of reflection of incoming plane-waves of energy $\omega^2$
in the potential (\ref{nowpot}).

Because our wavefunction with energy $\omega^2$ is not quite $A_x$, but is instead re-scaled by $f(\phi)$, there is a minor modification of
(\ref{condis}).  Following the simple logic in \S3\ of \cite{HorowitzRoberts}, we now find, for problems where the horizon is at $z=-\infty$, the asymptotic
region is at $z=0$, and the potential can formally be set to
vanish for $z > 0$, the formula:
\begin{equation}
\label{newcondis}
\sigma(\omega) = f^{2}(0)\left( {1-{\cal R} \over 1+{\cal R}} -i {f^\prime(0) \over f(0)
    \omega}~ \right).
\end{equation}
As expected, this reduces to (\ref{condis}) in the case of constant gauge-coupling, $f(\phi) = {\rm const.}$
In our case it follows from the equation of motion for the dilaton, eq.(\ref{third}) that   ${f^\prime(0)\over f(0)} \sim O(z^2)$
 in the asymptotic region.  Hence, 
 the second term in 
(\ref{newcondis}) will vanish sufficiently rapidly and drop out. Finally, choosing
$\phi_{0} $ such that $f^{2}=1$ asymptotically, we are left with the old formula.

We can make our Schr\"odinger problem a bit more concrete, and simplify the formulas, by using the equation of
motion satisfied by the background zeroth order gauge field $A_t$:
\begin{equation}
\label{atis}
\partial_r \left(f^2(\phi) r^2 e^{\chi \over 2} \partial_r A_t \right) = 0~.
\end{equation}
We find that the electrically charged black brane has
\begin{equation}
\partial_r A_t = {Q \over r^2} {1\over f^2(\phi)} e^{-{\chi \over 2}}~.
\end{equation}
Converting this to $\partial_z A_t$ and plugging in to (\ref{nowpot}),  the effective potential becomes
\begin{equation}
\label{finalpot}
V(z) = {1\over f} \left( f^{\prime\prime} + {ge^{-\chi} \over f}{Q^2\over r^4}\right)~.
\end{equation}

We can immediately see some interesting possibilities for different behaviors as compared to the cases studied in \cite{HorowitzRoberts}.
In those cases, the entire potential was proportional to $g$, which vanishes at the horizon, so it was guaranteed that there would be finite conductivity.
Here, that is not the case.
For instance, if $f \to 0$ at the horizon, which corresponds to ${\it strong}$ coupling, then the potential barrier may actually become
very large at the horizon, even though $g$ always has a zero there.  More controllably, a large value of ${f^{\prime\prime} \over f}$ at the horizon may
cause interesting behavior in the extremal limit.

We will now study this reflection problem for various choices of the dilaton coupling to the gauge field. 

\subsection{The canonical dilatonic black holes: $\frac{1}{4}f(\phi)^2 = e^{2 \alpha\phi}$}

To begin with, we should relate the metric gauge (\ref{newgauge}) to our old coordinates of \S2.1, to provide
the explicit form for the effective potential.  
Recall that in the $w$ coordinate of \S2, the extremal black brane has a metric:
\begin{equation}
ds^2 = -w^2 ~dt^2 + {dw^2\over w^2} + w^{2\beta}(dx^2 + dy^2)~.
\end{equation}
Matching metrics, we quickly find that
\begin{equation}
r = w^{\beta},~~g \sim r^2,~~e^{\chi} \sim r^{2(1-{1\over \beta})}~.
\end{equation}
Finally, from (\ref{zis}) we see that $z$ is related to $w$ by 
\begin{equation}
z = -\beta r^{-1/\beta} = -{\beta \over w}~.
\end{equation}

It is now easy to evaluate the effective potential (\ref{finalpot}) in this concrete case.  We find, using the solution for $e^{\alpha\phi}$ in \S2.1, that
in the near-horizon region:
\begin{equation}
V(z) = {c \over z^2}~.
\end{equation}
It further transpires that the constant $c$ is
${\it independent}$ of the value of $\alpha$:
\begin{equation}
\label{valcc}
c=2~.
\end{equation}
This is similar to the universality seen, in a different context, in \S5.3 of \cite{HorowitzRoberts}.  We will nevertheless continue our discussion with arbitrary values of $c$,
for possible use in related problems.

The change of variables from $z$ to $w$ takes the near-horizon region to $z=-\infty$ and asymptotia to $z = 0$.
Therefore, the scattering problem we wish to study has incoming plane waves with energy $\omega^2$ at $0$.

Because the potential has a ${1\over z^2}$ form, the WKB approximation does not apply.  
However, we can still solve this problem using the method of matched asymptotics, developed in roughly this context in \cite{GubserRoca}.
Define $\chi$ via
\begin{equation}
\label{chiis}
\Psi = \sqrt{-\pi \omega z \over 2}\chi~.
\end{equation}
Then the Schr\"odinger equation satisfied by $\Psi$ becomes
\begin{equation}
\label{chieq}
z^2 {\partial^2 \chi \over \partial z^2} + z {\partial \chi \over \partial z} + (z^2\omega^2 - \nu^2) \chi ~=0~,~~\nu^2 = c + {1\over 4}~.
\end{equation}

\subsubsection{Ingoing modes}

The AdS/CFT correspondence instructs us to choose modes which are purely ingoing at the horizon.
The solutions to the differential equation (\ref{chieq}) are Hankel functions; the ${\it purely~ingoing}$ solution, at the horizon, is given by
\begin{equation}
\chi = H_{\nu}^{(1)}(-\omega z) \sim \sqrt{{2\over -\pi \omega z}} {\rm exp} \left(-i\omega z - i (\nu + {1\over 2}) {\pi \over 2}\right)~
\end{equation}
where after $\sim$ we give the behavior as $z \to -\infty$.
Including time dependence, this yields a wavefunction
\begin{equation}
\label{psiis2}
\psi  \sim {\rm exp}\left(-i\omega (t+z) -i (\nu + {1\over 2}) {\pi \over 2}\right)~.
\end{equation}
This is the desired, purely ingoing, mode at the horizon.

\subsubsection{Overall strategy and detailed analysis}

We are now going to solve for the $\omega$-dependence of the resistivity by a method of matched asymptotics.  More precisely, we will do the following.
For small $\omega$, the $\omega^2$ term in the Schr\"odinger equation is really only relevant in the regions where the potential is arbitrarily small.  This happens
at $z = -\infty$ (the horizon) and at the boundary.  Therefore, {\it away} from these regions, it is surely a reasonable approximation to neglect the $\omega$-dependent
terms entirely, and use the approximation that $V(z)$ is the dominant term in the Schr\"odinger equation.  
We therefore start with the solution of the $\omega$-dominated equation near the boundary, and show that we can continue it (still using a near-boundary approximation)
to a solution in the potential-dominated region.  I.e., potential domination happens even very near the boundary, for sufficiently small $\omega$.  We then continue the
resulting solution all the way to the horizon, using appropriate matchings.  Finally, we use the conservation of flux between infinity and the horizon to determine the
$\omega$-dependence of the coefficient of the reflected wave.

\medskip
\noindent
{\bf Step 1: Near-boundary analysis}
\medskip

\noindent
For $r \gg r_{\rm horizon}$, we have
\begin{equation}
a^2 \sim r^2, ~~z \sim -{1\over r},~~V(z) \sim c_2 z^2~.
\end{equation}
We can therefore neglect the potential in the Schr\"odinger equation if we satisfy
\begin{equation} 
c_2 z^2 \ll \omega^2 \to r \gg {1\over \omega}~.
\end{equation}

This close to the AdS boundary, the equation has solutions
\begin{equation}
\label{boundarypsi}
\Psi(z) = D_1 e^{-i \omega z}~ +~ D_2 e^{i\omega z}~.
\end{equation}

Now, choose a point $z_1$ with $z_1^2 \ll \omega^2$.  It follows that in the small frequency limit, we also have
$|z_1| \ll {1\over \omega}$.    Therefore we can Taylor expand our wavefunction (\ref{boundarypsi}), yielding:

\begin{equation}
\Psi \sim (D_1 + D_2) + i \omega (-D_1 + D_2) z~.
\end{equation}

Now, still in the near-boundary region, we can choose a point $z_2$ where the ${\it potential}$ now dominates the $\omega^2$ term in
the Schr\"odinger equation.  This just requires $z_2^2 \gg \omega^2$, and is possible in the near-boundary region for small $\omega$.  Let us
suppose that while $V(z) \gg \omega^2$ in the vicinity of this point, in truth ${\it both}$ terms are negligible in the Schr\"odinger equation.  The conditions
for this to hold are that:
\begin{equation}
V(z) (\Delta z)^2 \ll 1,~~\omega^2 (\Delta z)^2 \ll 1,~~\Delta z \equiv \vert z_2 - z_1\vert~.
\end{equation}
These conditions can be satisfied for small frequency; they simply require that $\omega \ll |z_2| \ll 1$.

Then in reaching $z_2$ from $z_1$, we can neglect both potential and frequency terms in the Schr\"odinger equation, which means we can use linear extrapolation from 
$z_1$ to $z_2$!  Hence:
\begin{equation}
\label{finalbound}
\Psi \sim E_1 + E_2 z,
\end{equation}
and matching with (\ref{boundarypsi}) we find that
\begin{equation}
\label{values}
E_1 = D_1 + D_2,~~E_2 = i\omega (D_2 - D_1)~.
\end{equation}

\medskip
\noindent
{\bf{Step 2: Near-horizon analysis}}
\medskip

\noindent
We will momentarily try to match the wavefunction extrapolated from $z_2$, to a wavefunction in the near-horizon region.  What is the appropriate wave-function there?  For
$z$ approaching $-\infty$, as we have already discussed,
\begin{equation}
V(z) \sim {c \over z^2},~z \sim {-1\over {r-r_h}}~.
\end{equation}
We can find points in the near-horizon region where $V(z)$ dominates $\omega^2$; this simply requires $\vert \omega z\vert \ll 1$, and is true for arbitrarily large $|z|$ for small enough $\omega$.
Let us choose such a point, $z_3$.
In this region, $\vert \omega z_3\vert \ll 1$, the Hankel function reduces to
\begin{eqnarray}
\Psi \sim \sqrt{-{\pi \over 2}\omega z} H_{\nu}^{(1)}(-\omega z) & \sim &  \sqrt{-{\pi \over 2}\omega z} \left(J_{\nu}(-\omega z) + i N_{\nu}(-\omega z)\right) \nonumber \\
& \sim & \sqrt{-{\pi\over 2}\omega z} i {-(\nu -1)! \over \pi}
\left({-2 \over \omega z}\right)^{\nu}~. \label{mess}
\end{eqnarray}

\medskip
\noindent
{\bf{Step 3: Matching}}
\medskip

\noindent
Now, we need to match the wavefunction (\ref{finalbound}) with coefficients (\ref{values}) to the wavefunction (\ref{mess}).  This involves using the Schr\"odinger equation to integrate from
the point $z_2$ (near the boundary) to the point $z_3$ (near the horizon).  The key point, however, is that \textit{in the entire intermediate region, we can neglect the frequency dependence}
in the Schr\"odinger equation.  Therefore, the frequency dependence of $E_1$ and $E_2$ in (\ref{values}) can be determined from the frequency dependence we see in (\ref{mess}).
This yields
\begin{equation}
\label{theans}
E_1,~E_2 \sim \omega^{{1\over 2} - \nu} ~\to~ D_1 + D_2 \sim \omega^{{1\over 2}-\nu},~D_2 - D_1 \sim \omega^{-{1\over 2} - \nu }~.
\end{equation}

Next, how do we determine the conductivity $\sigma$?  The key point, as observed in \cite{HorowitzRoberts} following \cite{GubserRoca}, is that the exact Schr\"odinger equation has a conserved
flux
\begin{equation}
{\cal F} = i\left(\Psi^* \partial_z \Psi - \Psi \partial_z \Psi^*\right)~.
\end{equation}
Evaluating the frequency dependence close to the horizon, we find
\begin{equation}
{\cal F} \sim \omega~.
\end{equation}
Now at the boundary, we can write \cite{HorowitzRoberts}
\begin{equation}
{\cal F} \sim \vert D_1 + D_2 \vert^2 ~\omega \left({\rm Re} (\sigma)\right)~.
\end{equation}
This immediately fixes
\begin{equation}
\label{drumroll}
{\rm Re}(\sigma) \sim \omega^{2\nu - 1}~.
\end{equation}
Finally, noting that  for all values of $\alpha$, $\nu = 3/2$, we find
\begin{equation}
{\rm Re}(\sigma)\vert_{\rm dilaton ~black ~hole} \sim \omega^2~.
\end{equation}

The $\omega^2$ behavior of the conductivity, independent of the value of $\alpha$,   is intriguing. We do not have a good understanding
 for this universal result.   Mathematically it arises because the coefficient $c$ in the near-horizon potential always takes the value $2$, 
eq.(\ref{valcc}). 
However one gets the feeling that something deeper is at work here which  merits further understanding.
The same behavior of the conductivity was also obtained in some cases in \cite{HorowitzRoberts}, and has been proved to hold in general for
black branes with a near-horizon $AdS_2$ geometry in \cite{Leigh,Paulos}.  However, here we find that this behavior emerges under far more
general circumstances.

Let us also note that  the real part of the    conductivity  should have a delta function Drude peak at $\omega=0$.
This follows  on general grounds from the conservation of momentum. It is also a consequence of the Kramers Kronig relation and an 
expected  pole in the imaginary part of $\sigma$, see for e.g., \cite{HKMS}, \cite{Hartnoll}.  We have ignored this delta function contribution in the discussion above and 
have only focused on the  behavior 
of the remaining  regular part of the conductivity.

\subsection{More general attractors}
There is reason to believe that the universal low-frequency behaviour of the conductivity found above  
 continues to be true even for some of the other classes of attractors 
considered in section 2.2.1.  Here we present some additional evidence in support of this, leaving a more detailed analysis for the future.  
\subsubsection{Case 3}
Consider as an example case 3) which is a  limiting situation where the potential vanishes very  rapidly.
 In this case the effective potential takes the form
\begin{equation}
V_{eff}(\phi) = Q^2~ {\rm Exp}\left(-A e^{\alpha\phi} \right)~,
\end{equation}
which would arise in a theory where the gauge coupling function $f^2$ is characterized by
\begin{equation}
\frac{1}{4}f(\phi)^2 = {\rm Exp}\left(A e^{\alpha\phi} \right)~.
\end{equation}

Expressing everything in terms of $z$, and substituting the solution \eqref{exp5} into the effective Schr\"odinger potential (\ref{finalpot}), we find
\begin{equation}
\label{3pot}
V(z) = \frac{2}{z^2}\left(1 + \frac{3}{\log (-z)}+ \ldots \right)
\end{equation}

The  leading order potential  near the horizon, where $|z| \rightarrow \infty$, is therefore still $2/z^2$. As a result we expect the low frequency 
conductivity to behave as $\sigma \sim \omega^2$ in this case as well.

\section{Discussion}

In this paper, we have shown that charged dilaton black branes in AdS provide a rich
playground for studying black hole physics and holographic condensed matter physics.    The basic features, for the standard charged dilaton
branes with gauge-coupling function $f^2 \sim e^{2\alpha\phi}$, are:

\medskip
\noindent
$\bullet$
The near-horizon metric of the black holes has a Lifshitz-like symmetry in the metric, with a dynamical critical exponent $z$ that depends on $\alpha$, although the
full background solution breaks the symmetry.  This near-horizon structure is universal for black holes of arbitrary charge and asymptotic coupling, at fixed $\alpha$;
this represents a generalization of the attractor mechanism to this class of (much less symmetric) black branes.

\medskip
\noindent
$\bullet$
The ground state at finite charge density has vanishing entropy at extremality, and positive specific heat,
as expected for a garden-variety condensed matter system.

\medskip
\noindent
$\bullet$ The $T=0$ AC conductivity behaves (apart from a delta function at zero frequency)
 as $\sigma(\omega) \sim c(\alpha) ~\omega^2$ for all values of $\alpha$.\footnote{More correctly, as was discussed above,
 in view of possible corrections becoming important in the near-horizon region it is best to introduce a small 
non-zero temperature and interpret this result as the conductivity in the range, $ T\ll \omega \ll \mu$.} 
This universality is intriguing, and needs to be understood better.

\medskip
\noindent
$\bullet$
At finite but low temperature, there are plentiful low-energy degrees of freedom.  Compared to a 2+1 dimensional CFT where
the entropy density (and shear viscosity \cite{KSS}) scale like $T^2$ at low temperature, our system
has $s \sim T^{2\beta}$ (with a similar behavior for the viscosity).   Since $\beta < 1$, 
there are ${\it more}$ low-energy
degrees of freedom in these states with finite charge density than would be present in a CFT.   

\medskip
\noindent
$\bullet$
Our system, for $\beta = {1\over 2}$ ($\alpha = \sqrt{2}$), has similarities to a Fermi gas in the crudest thermodynamic properties.
However, we expect that the correlation functions actually  agree with those of a Lifshitz-like theory 
with $\beta={1\over 2}$. 
 
\medskip

Some clear directions for further work are as follows:

\medskip
\noindent
$\bullet$
  It will be interesting to compute the temperature dependence of the conductivity  for frequencies
in the range $\omega \ll T \ll  \mu$ and see how this scales with $T, \mu$.   

\medskip
\noindent
$\bullet$
Instead of the electrically charged case we can easily study a magnetically charged black brane using electromagnetic duality. 
In the dual CFT this gives us the response at non-zero magnetic field. Once again at zero temperature the entropy vanishes,
and the duality map allows the thermodynamic properties to be easily deduced. 

Perhaps more interesting is the case of a dyonic black brane, which carries both
 electric and magnetic charge. Here the effective potential
which governs attractor behavior has a  minimum which is at a finite value for the dilaton.
The resulting entropy is now non-zero and the near-horizon geometry is $AdS_2\times R^2$ as in the  extremal Reissner-Nordstr\"om black brane case. 
It will be interesting to see if the resulting thermodynamics and transport properties throw further light on the nature of charge carriers
and their interactions in the boundary field theory. 

\medskip
\noindent
$\bullet$
Our considerations can be easily generalized to include an axion. In this
 case the duality group can be promoted in supergravity to a full $SL(2,R)$ action,
using which  solutions where both the  axion and dilaton are activated and which carry both 
electric and magnetic charges can be obtained. The dual 
field theory would now have a non-zero Cherns Simons term  and this could lead to a rich set of possibilities both for critical behavior
and for possible insulator behavior
 obtained by deforming away from the critical point \footnote{We thank  Kedar Damle for 
discussions in this regard.}.

\medskip
\noindent
$\bullet$
Several recent papers have found interesting evidence of non-Fermi liquid states arising from the physics of probe fermions
in the AdS-RN black hole \cite{Rey,Sungsik,Johnone,Zaanen,Johntwo}.  It could be worthwhile to generalize these considerations to the family of solutions discussed here,
since the  thermodynamic properties are more amenable to a conventional dual interpretation.

We leave these fascinating questions for the future. 



\bigskip
\centerline{\bf{Acknowledgements}}
\medskip

S.K. would like to thank the participants of the ``Quantum Criticality and the AdS/CFT
correspondence'' miniprogram at KITP for inspirational discussions.  He is particularly grateful
to Mike Mulligan for repeatedly raising the question of the intriguing thermodynamic properties
of extremal black
holes.
He would also like to thank the faculty of the MIT Center for Theoretical Physics, particularly Krishna Rajagopal, for
interesting comments about this work during a seminar at MIT.  Other useful discussions with Allan Adams,  Tom Faulkner, Gary Horowitz,
 Andreas Karch, John McGreevy, Matt
Roberts,
Eva Silverstein and Sho Yaida are gratefully acknowledged.
S.K. was supported in part by the National Science Foundation under grant numbers
PHY05-51164 and
PHY-0244728, and the DOE under contract DE-AC03-76SF00515.

We would all like to acknowledge useful discussion with Nori Iizuka.

S.P.T.   would like to thank the SITP,  Stanford University, and SLAC for  their hospitality and support during his 
sabbatical visit when this research was begun.
He is deeply grateful to Mohit Randeria and Nandini Trivedi for several enlightening conversations and for sharing their lecture notes etc.
S.P. would like to thank R. Loganayagam for several informative discussions.
 S.P. and S.P.T. acknowledge  discussion
 with Sumit Das, Deepak Dhar, Sourendu Gupta,  Shiraz Minwalla, Gautam Mandal and  Spenta Wadia.
They are especially  indebted to their ``advisor'' Kedar Damle for  a 
much needed education in condensed matter physics! Most of all they thank the people of India for generously supporting 
research in string theory, and all the other subjects it is increasingly connected to.

\appendix
\section{Extremal branes: from near-horizon to infinity in AdS}
In this appendix, we obtain numerical solutions that interpolate between the near-horizon scaling solution eq.\eqref{attractora} and $AdS_4$.

The strategy for obtaining such a solution is to numerically integrate the equations of motion \eqref{first}, \eqref{second}, and \eqref{third} using standard techniques (NDSolve in Mathematica 7) starting near $w=0$ with initial data taken from the near-horizon solution.

However, the near-horizon solution is exact, so numerical integration using initial data drawn from it simply reproduces the near-horizon solution unmodified.  To numerically integrate to a solution that is asymptotically $AdS_4$, we must also take into account the subleading corrections to the near-horizon solution. (This is analogous to the case of an extremal Reissner-Nordstr\"om black hole -- the near horizon-solution is an exact solution to the equations of motion, however, subleading near-horizon corrections are permitted and allow the black hole to be embedded in asymptotically flat spacetime.) 

By adding the subleading correction, we introduce an additional parameter in the near-horizon solution -- the strength of the perturbation.

\subsection{Allowed corrections to the near horizon solution}
We must look for the allowed corrections to the near horizon solution eq.\eqref{attractora}.

We start with a fairly general ansatz for the modification to the metric:
\begin{eqnarray}
a(w)  & = & C_2 w \left(1 + d_1 w^{\nu_1} \right) \nonumber \\
 b(w) & = &  w^\beta \left(1 + d_2 w^{\nu_2} \right) \label{correctionsab}
\end{eqnarray}

The form of the perturbation of $\phi$ is determined from the ansatz for $b$ by the equation of motion \eqref{second}: 
\begin{equation}
  \phi(w) = -K\log (w) + C_3 + d_3 w^{\nu_2} \label{correctionsphi}
\end{equation}
where $d_3 = \frac{2\beta+\nu_2-1}{2 K} d_2$.


We first note that $\nu_1 = \nu_2$. This can be seen by substitution of the ansatz into eq.\eqref{first}. Since we require both $\nu_1$ and $\nu_2$ to be positive, the two terms proportional to $w^{\nu_1}$ and $w^{\nu_2}$ cannot separately cancel, so $\nu_1 = \nu_2 \equiv \nu$. (Even if we allow negative solutions, it turns out that allowing $\nu_1 \neq \nu_2$ yields only one consistent perturbation: $\nu_1 = \frac{-4+3\alpha^2}{4+\alpha^2}$ for which $d_2 = 0$. This solution can also be obtained considering the perturbation with $\nu_1 = \nu_2$.)

We now substitute the ansatz into eq.\eqref{first} and eq.\eqref{third}, which we solve to leading order in $w$. We will use one of the equations to solve for $d_1$ in terms of $d_2$ and $\nu$. Substituting into the remaining equation will result in a quartic equation for $\nu$. There will be no constraint on $d_2$, which is a free parameter that determines the strength of the perturbation. (The structure is similar to that of an eigenvalue problem: We are looking for vectors $(d_1 ~ d_2)$ in the kernel of some $2 \times 2$ matrix. The matrix depends on $\nu$ and $\nu^2$, hence we expect the condition that the determinant of the matrix vanishes to yield a quartic equation for $\nu$.)

Substituting the ansatz into eq.\eqref{first}  implies 
\begin{equation}
d_1 = \left( \frac{2(1+\beta)(1+2\beta)}{(2\beta+2+\nu)(2\beta+1+\nu)}-1 \right) d_2
\label{d1d2}
\end{equation}

Using this expression for $d_1$, eq.\eqref{third} is satisfied to leading order if $\nu$ satisfies the following quartic equation:
\begin{equation}
  (\nu +1) (4 \beta +\nu ) \left(-4 \beta ^2+(2 \beta +1)
   \nu -6 \beta +\nu ^2-2\right) =0
\end{equation}

The only positive root is 
\begin{eqnarray}
\nu & = & \frac{1}{2} \left(-2 \beta +\sqrt{(2 \beta +1) (10 \beta+9)}-1\right) \nonumber \\
 & = & \frac{-3 \alpha ^2+\sqrt{57 \alpha ^4+184 \alpha
   ^2+144}-4}{2 \left(\alpha ^2+4\right)}. \label{nuroot}
\end{eqnarray}
This agrees with eq.(\ref{expgamma}). 

To find all allowed perturbations we must also consider values of $\nu$ for which eq.\eqref{d1d2} is singular -- either $d_1=0$ or $d_2=0$. This happens if $(2\beta+2+\nu)(2\beta+1+\nu)=0$. Both these roots are negative, so they do not concern us here. However, for reference we note that the finite temperature solution eq.\eqref{temp} is obtained from choosing $\nu=-2\beta-1=-\frac{3 \alpha ^2+4}{\alpha ^2+4}$ for which $d_2=0$. We have not explored what happens when we consider the other negative values of $\nu$, perhaps they also give rise to interesting solutions.

Finally, we observe that the constraint eq.\eqref{constraint} is satisfied for $\nu$ given by eq.\eqref{nuroot} and $d_1$ given by eq.\eqref{d1d2}, as required. 

The final form of the perturbed solution is eq.\eqref{correctionsab} and eq.\eqref{correctionsphi} with $\nu$ given by eq.\eqref{nuroot}, and $d_1$ related to $d_2$ according to eq.\eqref{d1d2}.

\subsection{Numerical integration}
Here we present results of the numerical integration. 

\begin{figure}
  \begin{center}
  \includegraphics{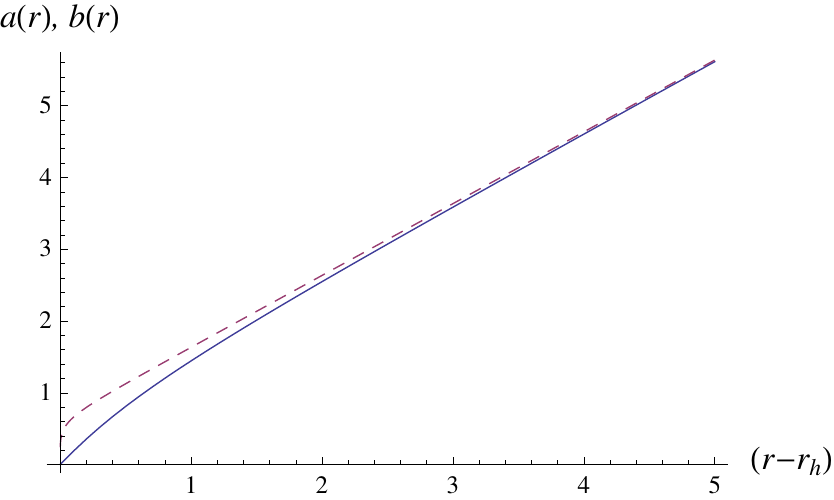}

  \includegraphics{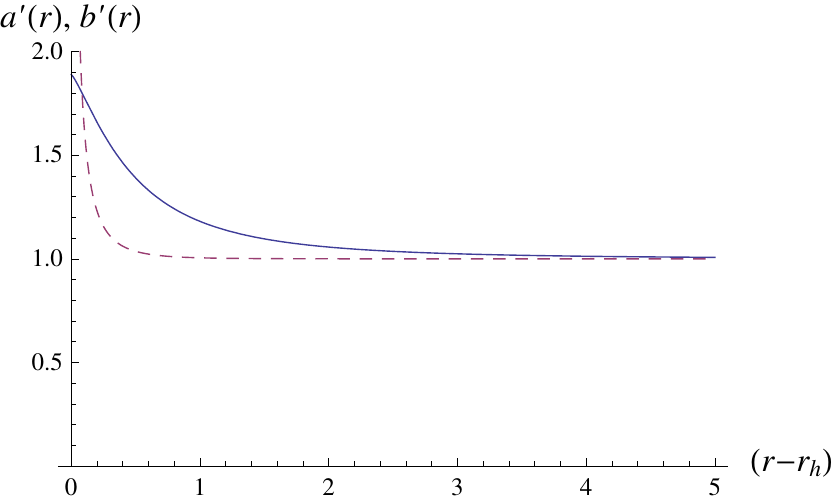}
  \end{center}
  \caption{ \label{plotA} Numerical solution interpolating between the near horizon solution and $AdS_4$ for $\alpha=1$ and $d_1 =-.514219$. The second plot shows that $a'(r)$ and $b'(r)$ approach $1$. Solid lines denote $a$, dashed lines denote $b$.} 
\end{figure}

\begin{figure}
\begin{center}
\includegraphics{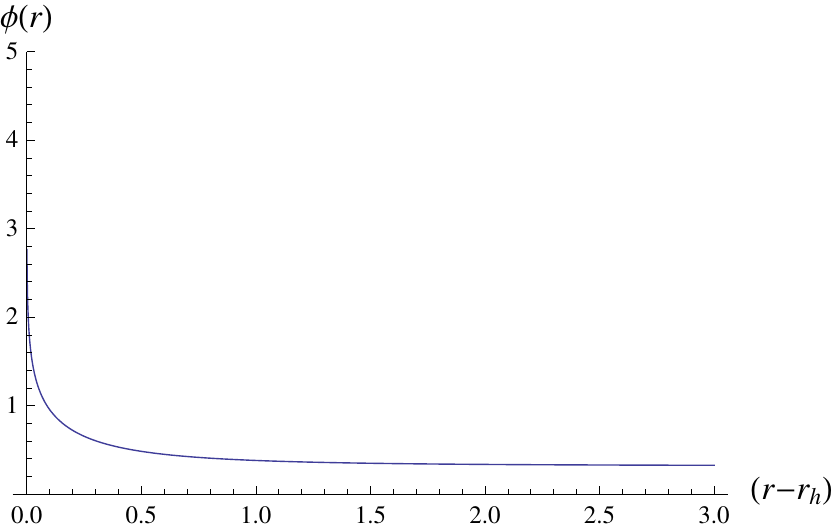}
\end{center}
\caption{\label{plotPhi} Numerical solution for $\phi$, for $\alpha=1$.}
\end{figure}

Initial data for numerical integration is taken from the modified near-horizon solutions \eqref{correctionsab} and \eqref{correctionsphi}. Figures \ref{plotA} and \ref{plotPhi} show the resulting solution for $\alpha=1$. The strength of the perturbation was
chosen to be  $d_1 =-.514219$, so that the solution meets the  condition that $b'(r) \rightarrow 1$ as $r \rightarrow \infty$. 
(For other negative values of $d_1$, $b'(r)$ approaches a constant which is different from one,  a coordinate transformation then brings 
the solution back to  a form with the standard asymptotics of $AdS_4$ space. For  positive $d_1$ the numerical solution becomes singular.)

Figures \ref{plotA} and \ref{plotPhi} clearly show that $a(r) = r$ and $b(r) = r$ for large $r$, so the solution is 
asymptotically $AdS_4$. The dilaton approaches a constant, $\phi_0$. 

A similar solution is also obtained for other values of $\alpha$.
 
The solution  above corresponds to a particular value of the two parameters $Q$ and $\phi_0$, determined by the choice of 
gauge for the near-horizon solution \eqref{attractora}. However, it is straightforward to use a scaling symmetry and the 
freedom to add a constant to $\phi(r)$ to obtain solutions for any values of $Q e^{-\alpha \phi_0}$ and $\phi_0$. We can obtain a valid solution with a different value of $\phi_0$ by adding a constant $\delta\phi_0$ to the solution $\phi(r)$; with $Q e^{-\alpha \phi_0}$ unchanged. Solutions for different values of $Q^2 e^{-2\alpha \phi_0}$ can be obtained by a rescaling  $r \rightarrow \lambda r$, $t \rightarrow t/\lambda $ and $x_i \rightarrow x_i/\lambda$ as described in the main text.




\end{document}